\RequirePackage[2020-02-02]{latexrelease}
\documentclass[twocolumn,twoside]{revtex4}
\usepackage{graphicx}
\usepackage{fancyhdr}
\usepackage[T1]{fontenc}
\usepackage{relsize}
\def\babar{\mbox{\slshape B\kern-0.1em{\smaller A}\kern-0.1em
		B\kern-0.1em{\smaller A\kern-0.2em R}}}
\begin{document}

\title{Measurement of Beam Polarization at an $e^+e^-$ B-Factory with New Tau Polarimetry Technique}

%

\author{Caleb~Miller\\
on behalf of the \babar~collaboration}
\affiliation{University of Victoria, Victoria, BC, Canada}

\begin{abstract}
The \babar~collaboration has demonstrated the first application of the new Tau Polarimetry technique. This polarimetry technique exploits the kinematic coupling of $\tau$ decay products to the spin states of the $\tau$ and initial state electron, to precisely determine the average beam polarization in an $e^+e^-$ collider. The Tau Polarimetry technique is expected to be used at Belle II following an upgrade to a polarized electron beam, where the precision with which the average beam polarization is known is expected to be the dominant systematic uncertainty in proposed future measurements which require polarized beams. Applying Tau Polarimetry, \babar~has preliminarily measured the PEP-II average beam polarization to be $\langle P\rangle=-0.0010\pm0.0036_{\textrm{stat}}\pm0.0030_{\textrm{sys}}$.
\end{abstract}

\maketitle

\thispagestyle{fancy}


\section{Introduction}
Belle II has proposed an upgrade for SuperKEKB to introduce a polarized electron beam\cite{PolarizationWhitePaper}. This proposed upgrade would allow for a number of world-leading precision measurements in the electroweak sector. It is expected that for these proposed measurements the dominant systematic uncertainty will be the precision with which the average beam polarization is known. As such \babar~has developed and demonstrated a novel technique for extracting the beam polarization directly from data collected within the detector. This technique exploits two key features present in $e^+e^-\rightarrow\tau^+\tau^-$ events. The first key feature is the coupling between the beam polarization and the polarization of the $\tau$ produced, which at a center-of-mass energy of 10.58 GeV can be expressed as:
\begin{eqnarray}
	P_\tau & = & P_e\frac{cos\theta}{1+cos^2\theta} \\
	&&-\frac{8G_Fs}{4\sqrt{2}\pi\alpha}g^\tau_V\left(g^\tau_A\frac{|\vec{p}|}{p^0}+2g^e_A\frac{cos\theta}{1+cos^2\theta}\right)  \nonumber
	\label{eqn:ePoltoTauPol}
\end{eqnarray}
The second key feature exploited in the Tau Polarimetry technique is the coupling of the $\tau$ polarization to the kinematics of the final state particles produced in the $\tau$ decay. This is most easily conceptualized in the $\tau^\pm\rightarrow\pi^\pm\nu_\tau$ decay. In this decay the single chiral state available to neutrino forces it to be emitted forward or backward relative to the $\tau$ momentum vector. ie. For a left-handed $\tau^-$ the neutrino must have momentum parallel to the $\tau$ direction. In turn, due to the two-body decay, this restricts the pion to be emitted anti-parallel to the $\tau$ direction and therefore will have a lower average momentum. This momentum dependence then flips to higher on average for a right-handed $\tau^-$. Combining these key features with a sufficiently large $\tau$ dataset allows the average beam polarization to be measured.

\section{Event Selection}
This measurement uses 424 fb$^{-1}$ of $\Upsilon(4S)$ data collected at the \babar~detector\cite{babar,babar2}. Of the available 424 fb$^{-1}$, 32.28 fb$^{-1}$ were used to study and develop the analysis and are excluded from the final result. Therefore the final measurement is preformed with 392 fb$^{-1}$ of data. In developing the analysis it was found that the  hadronic $\tau$ decays contain the most beam polarization sensitivity, and requiring the second $\tau$ to decay leptonically provides a low background environment. For this analysis the $\tau^\pm\rightarrow(\rho\rightarrow\pi^\pm\pi^0)\nu_\tau$ decay was chosen as the signal due to the large branching fraction. The second $\tau$ is required to decay via an electron, $\tau^\pm\rightarrow e^\pm\nu_\tau\overline{\nu}_e$, which results in a 99.7\% pure $\tau$-pair sample. In order to select for this $e$-$\rho$ decay topology, the events are required to contain two charged particles which are in opposing hemispheres as defined by the thrust axis in the center-of-mass frame. One of the charged particles must be tagged as an electron by the \babar~particle identification algorithms, and have no neutral particles identified in its hemisphere. The signal $\tau$, the rho decay, is required to have a neutral pion in its hemisphere either identified by the \babar~particle identification algorithms or from a pair of neutral particles which have an invariant mass consistent with a neutral pion. Finally a requirement that the event exceeds 1.2 GeV of transverse momentum removes the majority of any remaining Bhabha and two-photon events.
\section{Polarization Measurement}
In order to extract the beam polarization from the data sample a Barlow and Beeston template fit is implemented\cite{Barlow}. For the $\rho$-decay mode, the fit requires three variables to extract the average beam polarization. First, $\cos\theta$ is required due to the sign of $P_\tau$ mirroring the sign of $\cos\theta$ as seen in Equation \ref{eqn:ePoltoTauPol}. Next, due to the spin 1 nature of $\rho$, two variables are required to disentangle the polarization information. $\cos\theta^*$, which is the angle between the $\tau$ and $\rho$ momentum vectors in the $\tau$ rest frame, and $\cos\psi$, which is the angle between the $\rho$ and the pion momentum vectors in $\rho$ rest frame. These 3 angular variables are binned into 3D histograms for the fitting algorithm. The electric charge of the $\tau$ is also important for the fit as the polarization coupling flips sign with electric charge as well. This effect is accounted for by performing the fit independently for the two charge states, and then combining the two fits to arrive at the final result. In order to be sensitive to the beam polarization, $\tau$ Monte Carlo (MC) was produced with {\tt KKMC}\cite{kkmc} for both a left and right handed beam polarization. The fit then uses both of these polarized MC samples along with existing \babar~non-$\tau$ MC as templates in the fit. The average beam polarization is defined as the difference between the fitted contribution for the left and right polarized templates. For the \babar~data an average beam polarization of -0.0011$\pm$0.0036 was found.
\section{Polarization Sensitivity}
\begin{figure}
	\includegraphics[width=\linewidth]{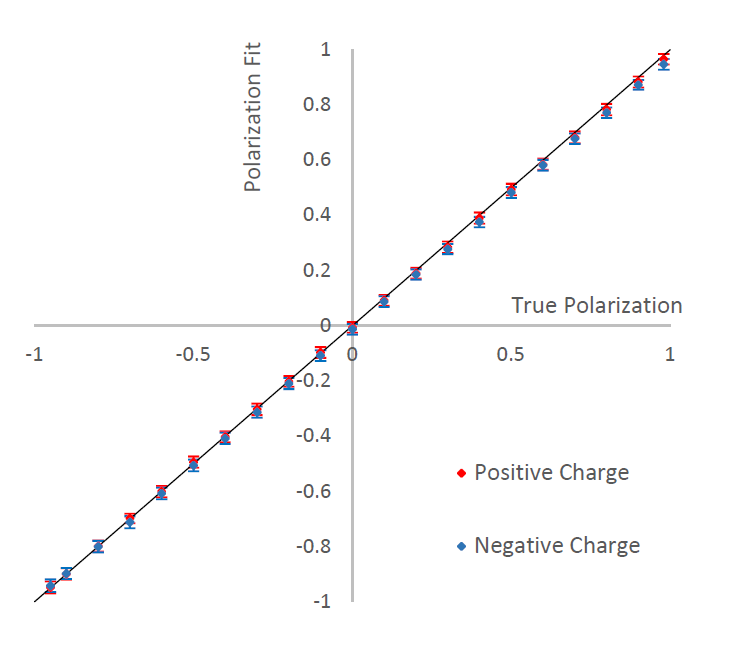}
	\caption{Beam polarization measured in MC samples as a function of input beam polarization.}
	\label{fig:polarsense}
\end{figure}

Since PEP-II is expected to have no inherent beam polarization, \babar~has carried out MC studies of the Tau Polarimetry performance at arbitrary beam polarization states. This is performed by splitting the generated polarized $\tau$ MC into two samples. One to mix a specified beam polarization state, and the other to perform the polarization fit. Figure \ref{fig:polarsense} shows the response of the fit to various input polarizations, and demonstrates that Tau Polarimetry performs well at any beam polarization state.
\section{Systematic Uncertainties}
The systematic uncertainties were evaluated by quantifying the shift in the level of agreement between data and MC fits in response to a systematic shift in an analysis variable. Systematic uncertainties were then evaluated independently for each of the 6 \babar~datasets (Runs 1-6). Each systematic source is then combined across runs in a way that accounts for correlations in the uncertainties. Finally all the systematic uncertainties are summed in quadrature to arrive at a final systematic uncertainty. The dominant systematic uncertainty in this analysis arose from uncertainties associated with the \babar~neutral pion identification algorithms, which contribute an uncertainty of 0.0015. The sub-leading uncertainties arise from the modelling of neutral split-offs from charged particles(0.0011), modelling of $\cos\psi$(0.0010), modelling of $\cos\theta$(0.0009), minimum neutral particle energy to be included in the event(0.0009), and neutral pion mass reconstruction(0.0009). When the systematic uncertainties are summed in quadrature a total systematic uncertainty of 0.0030 is found.
\section{Conclusions}
The \babar~collaboration has successfully implemented the first use of the novel Tau Polarimetry technique, and measured the PEP-II average beam polarization to be -0.0011$\pm0.0036_{\textrm{stat}}\pm0.0030_{\textrm{sys}}$. While the specific details of the systematic uncertainties will vary by detector we expect an upgraded Belle-II/SuperKEKB will be able to match or outperform the precision of this measurement. This level of precision exceeds the performance required for the physics projections in the Belle polarization upgrade proposal\cite{PolarizationWhitePaper} and as such further motivates the addition of beam polarization. It is also expected that Tau Polarimetry could be implemented at future $e^+e^-$ colliders which implement beam polarization, though a detailed study at the specific beam energy will be required.

\bigskip 

\end{document}